\documentstyle[aps,prb,multicol,epsfig]{revtex}
\begin{document}
% \draft command makes pacs numbers print
\draft
\title{
Localization length and impurity dielectric 
susceptibility 
in the critical regime\\ 
of the metal-insulator transition 
in homogeneously doped $p$-type Ge 
}
% repeat the \author\address pair as needed
\author{Michio Watanabe,$^1$ Kohei M. Itoh,$^{1,2}$ Youiti Ootuka$^3$, 
and Eugene E. Haller$^4$} 
\address{$^1$Department of Applied Physics and Physico-Informatics, 
Keio University, Yokohama 223-8522, Japan\\  
$^2$PRESTO-JST, Yokohama 223-8522, Japan\\  %3-14-1 Hiyoshi, Kohoku-ku, 
$^3$Institute of Physics, University of Tsukuba, 
1-1-1 Tennodai, Tsukuba, Ibaraki 305-8571, Japan\\  
$^4$Lawrence Berkeley National Laboratory and University of
California at Berkeley, Berkeley, California 94720}
\date{Received 10 April 2000}
\maketitle
\begin{abstract}
% insert abstract here
We have determined the localization length $\xi$
and the impurity dielectric susceptibility $\chi_{\rm imp}$ 
as a function of Ga acceptor concentrations ($N$) in nominally 
uncompensated $^{70}$Ge:Ga just below the critical concentration 
($N_c$) for the metal-insulator transition.  
Both $\xi$ and $\chi_{\rm imp}$ diverge at $N_c$ 
according to the functions $\xi\propto(1-N/N_c)^{-\nu}$ 
and $\chi_{\rm imp}\propto(N_c/N-1)^{-\zeta}$, respectively, 
with $\nu=1.2\pm0.3$ and $\zeta=2.3\pm0.6$ for $0.99N_c< N< N_c$.
Outside of this region ($N<0.99N_c$), the values of the exponents 
drop to $\nu=0.33\pm0.03$ and $\zeta=0.62\pm0.05$.  
The effect of the small amount of compensating dopants that 
are present in our nominally uncompensated samples, 
may be responsible for the change of the critical exponents at 
$N\approx0.99N_c$.  
\end{abstract}
% insert suggested PACS numbers in braces on next line
%\vspace*{5mm}

\begin{multicols}{2}

% body of paper here
The metal-insulator transition (MIT) in doped semiconductors is 
a unique quantum phase transition in the sense that both disorder 
and electron-electron interaction play a key role.~\cite{review}  
Important information about the MIT 
is provided by the values of the critical exponents 
for the zero-temperature conductivity, correlation length, 
localization length, and impurity dielectric susceptibility.  
From a theoretical point of view, the correlation length in 
the metallic phase and the localization length 
in the insulating phase diverge at the critical point with the 
same exponent $\nu$, i.e., they  are proportional 
to $|N/N_c-1|^{-\nu}$ in the critical regime of the MIT.  
($N$ is the dopant concentration 
and $N_c$ is the critical concentration for the MIT.)  
Since direct experimental determination of $\nu$ is extremely difficult, 
researchers have usually determined, 
instead of $\nu$, the value of 
$\mu$ defined by $\sigma(0)\propto (N/N_c-1)^{\mu}$ 
where $\sigma(0)$ is the conductivity extrapolated 
to $T=0$.~\cite{Ito96,Wat98}  
It is also possible to evaluate $\mu$ from finite-temperature scaling of 
the form $\sigma(N,T)\propto T^x f(|N/N_c-1|/T^y)$ where $x/y$ 
is equivalent to $\mu$.~\cite{Bog99,Waf99,Ito99}  
Values of $\nu$ are then obtained assuming $\nu=\mu$ 
for three-dimensional systems.~\cite{Weg76}  

In this work we have determined directly the localization 
length $\xi$ and the impurity dielectric susceptibility $\chi_{\rm imp}$ 
in neutron-transmutation-doped (NTD), 
nominally uncompensated $^{70}$Ge:Ga just below $N_c$. 
The application of NTD to isotopically enriched $^{70}$Ge 
leads to unsurpassed doping homogeneity and precisely controlled 
doping concentration. As a result, we have been able to 
approach the transition as close as $0.999N_c$ 
from the insulating side and $1.0004N_c$ from the metallic side.~\cite{Wat98}  
In zero magnetic field, the low-temperature resistivity of the samples is described 
by variable-range hopping (VRH) conduction within the Coulomb gap.~\cite{Shk84}  
Magnetic field and temperature dependence of the resistivity 
are subsequently measured in order to determine directly $\xi$ and $\chi_{\rm imp}$ 
in the context of the VRH theory.~\cite{Shk84}  

This kind of determination of $\xi$ and $\chi_{\rm imp}$ was 
performed for compensated Ge:As by Ionov {\em et al.}~\cite{Ion83}  
They found $\xi\propto(1-N/N_c)^{-\nu}$ 
and $\chi_{\rm imp}\propto(N_c/N-1)^{-\zeta}$
with $\nu=0.60\pm0.04$ and $\zeta=1.38\pm0.07$, respectively, for $N<0.96N_c$.  
The significance of their result is the experimental verification 
of the relation $2\nu\approx\zeta$ that had been predicted 
by scaling theories.~\cite{Kaw84,Imr81}  
However, the critical exponents of compensated samples are known to be 
different from those of nominally uncompensated samples. 
Therefore, the present work which probes $\xi$ and $\chi_{\rm imp}$ 
in nominally uncompensated samples is relevant for the 
fundamental understanding of the MIT. 
The previous effort to measure $\chi_{\rm imp}$ 
has also contributed.~\cite{Cas80,Hes82,Kat90}   Hess {\em et al.} 
found $\zeta=1.15\pm0.15$ in nominally uncompensated 
Si:P.~\cite{Hes82}  Since $\mu\approx0.5$ was determined for the same 
series of Si:P samples, $2\mu\approx\zeta$ was  again valid.  Katsumoto 
has found $\zeta\approx2$ and $\mu\approx1$ for compensated 
Al$_x$Ga$_{1-x}$As:Si, i.e., again, $2\mu\approx\zeta$ applies.~\cite{Kat90}  
Thus, in these cases the conclusion $2\nu\approx\zeta$ was reached indirectly, 
{\em by assuming} $\mu=\nu$. 
The work reported here, on the other hand, determines $\nu$ directly, i.e., 
we do not have to rely on the assumption $\mu=\nu$ in order to study the behavior 
of $\xi$ near $N_c$.  

All of the $^{70}$Ge:Ga samples used in this study 
were prepared by NTD of isotopically enriched $^{70}$Ge single crystals.  
We use the NTD process since it is known to produce 
the most homogeneous dopant distribution.~\cite{Ito96}  
Details of the sample preparation and characterization 
are described elsewhere.~\cite{Wat98}  
In this study, we determined the low-temperature ($0.05-0.5$~K) resistivity 
of nine samples in weak magnetic fields ($< 0.4$~T) applied 
in the direction perpendicular to the current flow.  

The electrical conduction of doped 
semiconductors on the insulating side of the MIT is often 
dominated by VRH at low temperatures.  
The temperature dependence of the resistivity $\rho (T)$ 
for VRH is written in the form of 
\begin{equation}
\label{eq:VRH}
\rho (T) = \rho_0(T) \exp [(T_0/T)^p\,],  
\end{equation}
where $p=1/2$ for the excitation within a parabolic-shaped energy gap 
(the Coulomb gap),~\cite{Shk84} 
and $p=1/4$ for a constant density of states around the Fermi level.~\cite{Mot90}  
In our earlier work,~\cite{Wat98} we reported that $p=1/2$ for $N<0.991N_c$ 
($N_c=1.860\times10^{17}$~cm$^{-3}$) and that $p$ decreases rapidly 
as $N$ approaches $N_c$ from $0.991N_c$ and becomes even smaller than 1/4 
when we neglect the temperature variation of $\rho_0(T)$.  
However, the variation 
contributes greatly to the temperature dependence of $\rho(T)$ near $N_c$ 
because the factor $T_0/T$ in the exponential terms become very small, i.e.,  
the temperature dependencies of $\rho_0(T)$ and that of the exponential term 
become comparable.  
Theoretically, $\rho_0(T)$ is expected to vary as $\rho_0 \propto T^{-r}$ 
but the value of $r$ including the sign has not 
been derived yet for VRH 
with both $p=1/2$~(Ref.~\ref{Shk98}) and $p=1/4$~(Ref.~\ref{All72}).  
%%%%%%%%%%%%%%%%%%%%%%%%%%%%%%%%%%%%%%%%%%%%%
%%%%  fig:T-dep                 %%%%%%%%%%%%%
%%%%%%%%%%%%%%%%%%%%%%%%%%%%%%%%%%%%%%%%%%%%%
\begin{figure}
\centerline{
\psfig{file=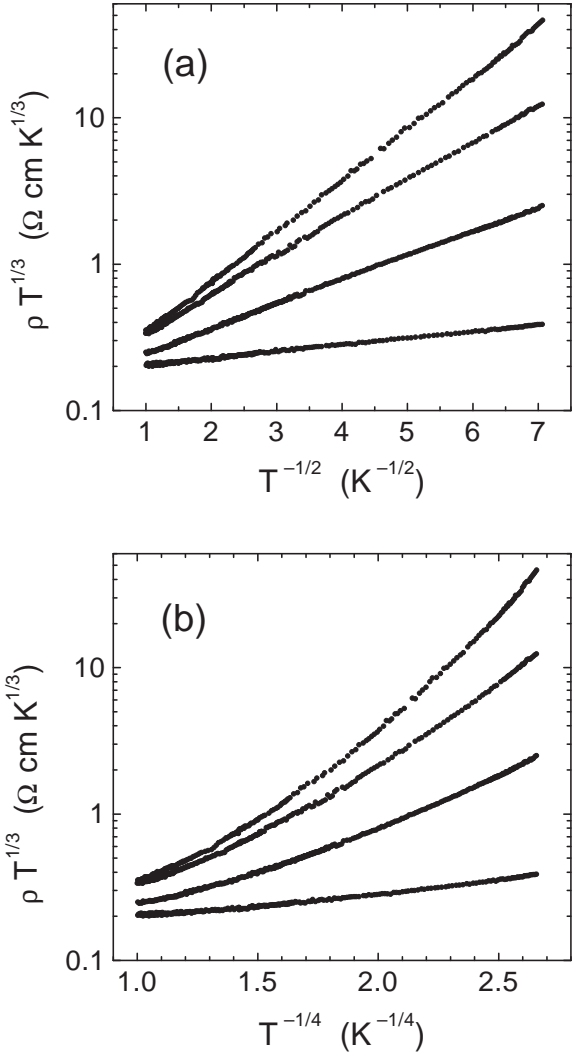,width=0.9\columnwidth,
bbllx=75,
bblly=65,
bburx=475,
bbury=800,clip=,
angle=0}
}
\caption{Resistivity %$\rho$ 
multiplied by $T^{1/3}$ 
vs (a) $T^{-1/2}$ and (b) $T^{-1/4}$  
for $^{70}$Ge:Ga.  From top to bottom in units of 10$^{17}$~cm$^{-3}$, 
the Ga concentrations are 1.848, 1.850, 1.853, and 1.856.}  
\label{fig:T-dep}
\end{figure}
%%%%%%%%%%%%%%%%%%

Recently, we have shown that the temperature dependence of the conductivity 
of the same series of $^{70}$Ge:Ga samples within $\pm$0.3\% of $N_c$ 
is proportional to $T^{1/3}$ at $0.02-1$~K.~\cite{Wat98}  
Since both the $T^{1/3}$ dependence of the conductivity and 
the Efros-Shklovskii VRH are results of the electron-electron
interaction in disordered systems, 
they can be expressed, in principle, in a unified form.
Moreover, the electronic transport in barely metallic samples and 
that in barely insulating samples 
should be essentially the same at high temperatures so long as 
the inelastic scattering length and the thermal diffusion length 
are smaller than, or at most comparable to  
the correlation length or the localization length.  
So, the temperature dependence of conductivity at high temperatures 
should be the same on both sides of the transition. Such behavior is 
confirmed experimentally in the present system,~\cite{Wat98} 
i.e., the conductivity of samples very close to $N_c$ 
shows a $T^{1/3}$ dependence at $T\approx0.5$~K, irrespective of the phase 
(metal or insulator) to which they belong at $T=0$.
Based on this consideration we fix $r=1/3$.
Figure~\ref{fig:T-dep} shows $\rho T^{\,r}$ with $r=1/3$ 
for four samples ($N/N_c=0.993$, 0.994, 0.996, and 0.998) 
as a function of (a) $T^{-1/2}$ and (b) $T^{-1/4}$.  
All the data points lie on straight lines with $p=1/2$ in 
Fig.~\ref{fig:T-dep}(a) while they curve upward with $p=1/4$ 
in Fig.~\ref{fig:T-dep}(b).  
This dependence is maintained even when we change the values of $r$ 
between 1/2 and 1/4.  Thus we conclude that the 
resistivity of all samples for $N$ up to 0.998$N_c$ is described 
by the VRH theory where the excitation occurs within the Coulomb gap, 
i.e., Eq.~(\ref{eq:VRH}) with $p=1/2$.

\begin{figure}
\centerline{
\psfig{file=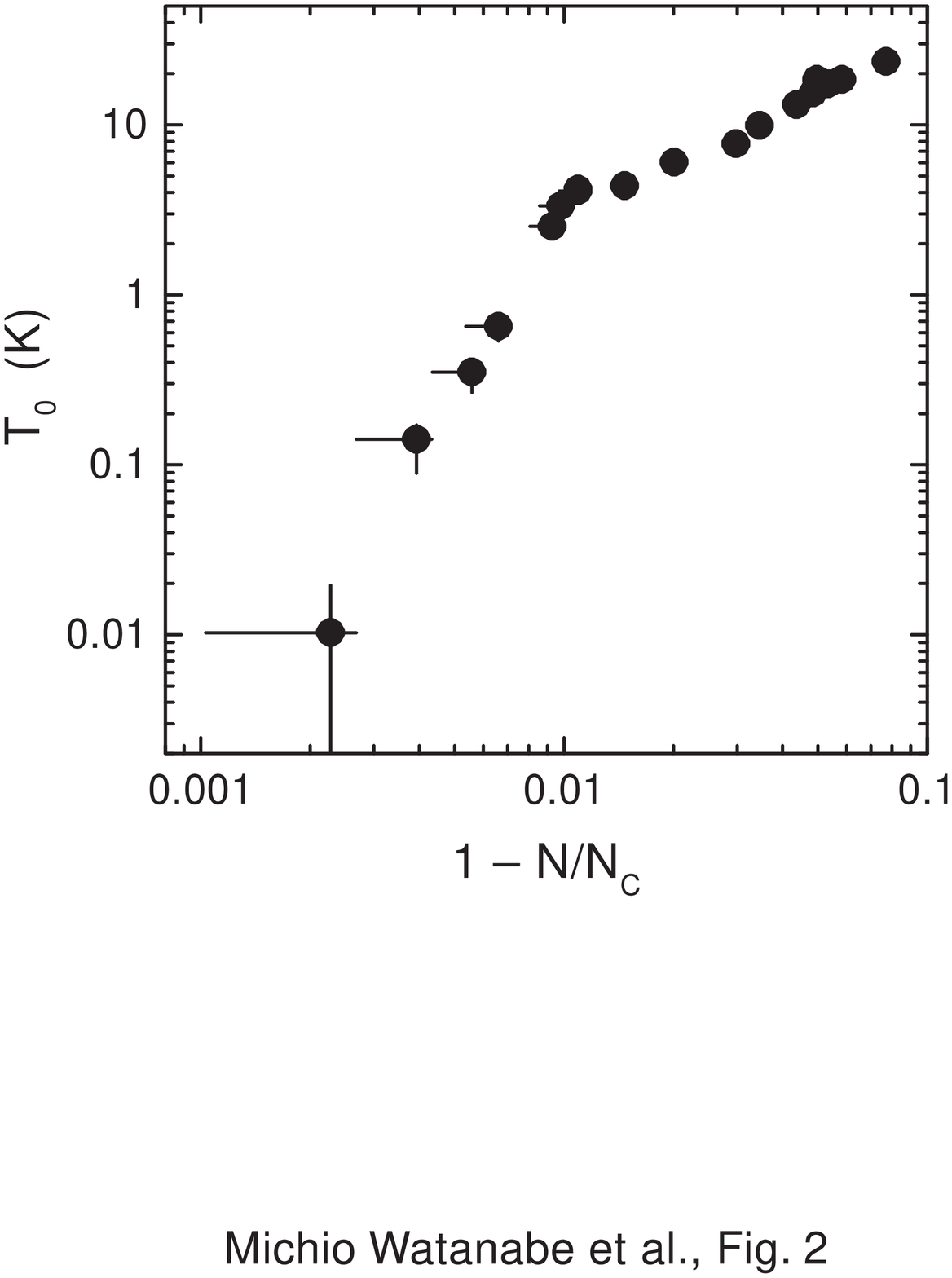,width=0.9\columnwidth,
bbllx=20,
bblly=245,
bburx=525,
bbury=720,clip=,
angle=0}
}
\caption{$T_0$ determined by $\rho(T) \propto 
T^{-1/3}\exp[(T_0/T)^{1/2}]$ as a function of 
the dimensionless concentration $1-N/N_c$.}  
\label{fig:T0}
\end{figure}

\begin{figure}
\centerline{
\psfig{file=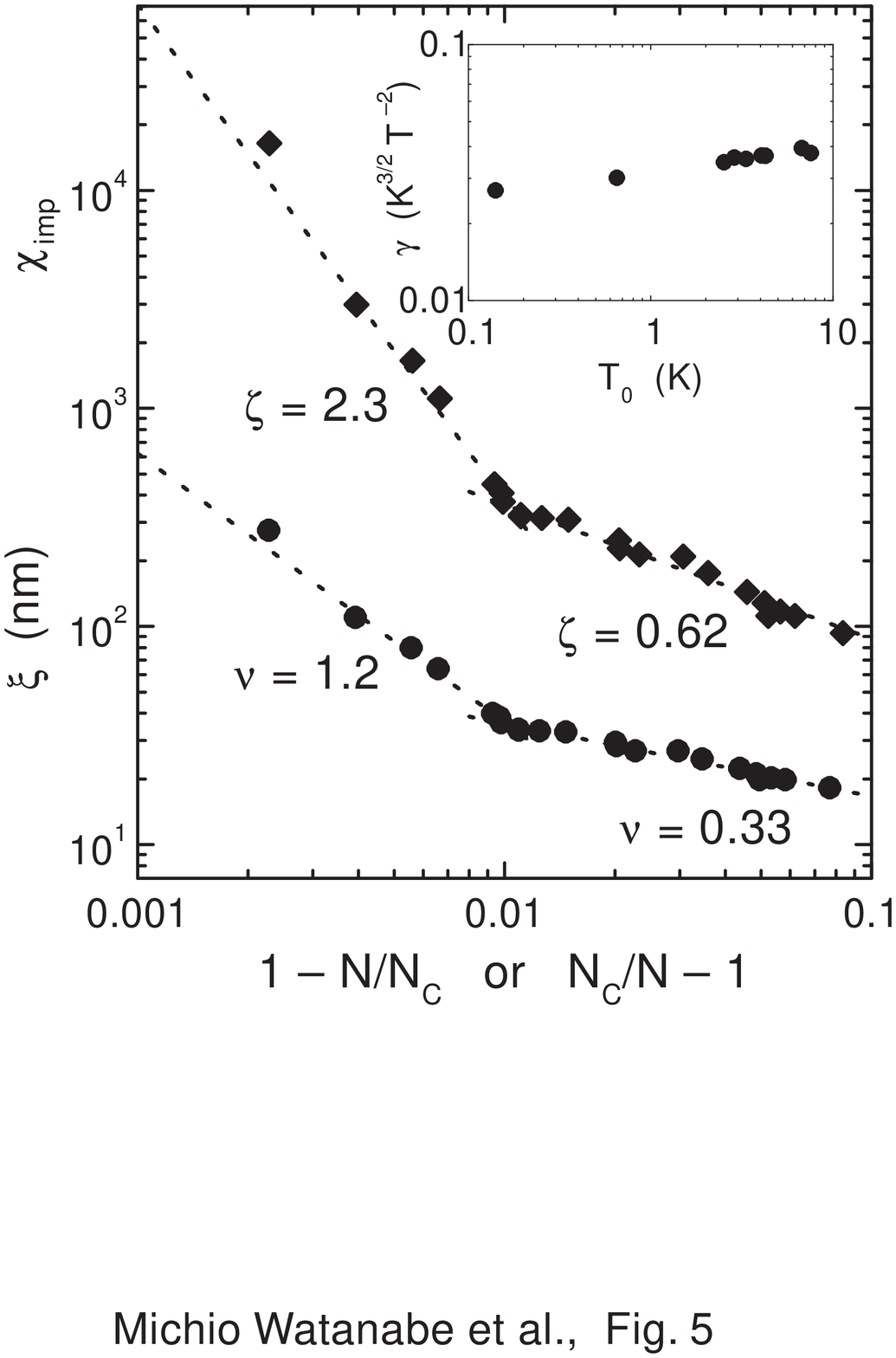,width=0.9\columnwidth,
bbllx=55,
bblly=220,
bburx=530,
bbury=765,clip=,
angle=0}
}
\caption{Localization length $\xi$ vs $1-N/N_c$ (lower data set) 
and the dielectric susceptibility $\chi_{\rm imp}$ arising 
from the impurities vs $N_c/N-1$ (upper data set).  
The inset shows the coefficient $\gamma$ defined 
by Eq.~(\protect{\ref{eq:gamma}}) as a function of $T_0$.  }
\label{fig:chixi}
\end{figure}

Based on these findings, we evaluate $T_0$ in 
Eq.~(\ref{eq:VRH}) with $p=1/2$ and $r=1/3$, 
and show it as a function of $1-N/N_c$ in Fig.~\ref{fig:T0}.  
The vertical and horizontal error bars have been estimated 
based on the values of $T_0$ obtained with $r=1/2$ and $r=1/4$, 
and the values of $1-N/(1.858\times10^{17}$~cm$^{-3})$ and $1-N/ 
(1.861\times10^{17}$~cm$^{-3})$, where $1.858\times10^{17}$~cm$^{-3}$ 
is the highest concentration in the insulating phase and 
$1.861\times10^{17}$~cm$^{-3}$ is the lowest in the metallic phase, 
respectively.  
%%%%%%%%%%%%%%%%%%%%%%%%%%%%%%%%%%%%%%%%%%%%%%%
%%%%%%%%  fig:T0  %%%%%%%%%%%%%%%%%%%%%%%%%%%%%
%%%%%%%%%%%%%%%%%%%%%%%%%%%%%%%%%%%%%%%%%%%%%%%
%%%%%%%%%%%%%%%%%%%%%%%%%%%%
According to theory,~\cite{Shk84} $T_0$ in Eq.~(\ref{eq:VRH}) 
is given by 
\begin{equation}
\label{eq:T0}
k_B T_0 \approx 2.8 e^2/4\pi\epsilon_0\,\epsilon(N)\,\xi(N)
\end{equation}
in SI units, where $\epsilon (N)$ 
is the dielectric constant. 
Here, we should note that the condition $T<T_0$ is needed 
for the theory to be valid, i.e., $T_0$ has to be evaluated only from the data 
obtained at temperatures low enough to satisfy the condition.  
This requirement is fulfilled in Fig.~\ref{fig:T0} 
for all the samples except for the one with $N=0.998N_c$.  
Concerning this latter sample, we will include it for the determination 
of $\xi$ and $\chi_{\rm imp}$ (Fig.~\ref{fig:chixi}) 
but not for the calculation of the critical exponents.  

%%%%%%%%%%%%%%%%%%%%%%%%%%%%%%%%%%%%%%%%%%%%
%%%%%%  fig:B-dep  %%%%%%%%%%%%%%%%%%%%%%%%%
%%%%%%%%%%%%%%%%%%%%%%%%%%%%%%%%%%%%%%%%%%%%
\begin{figure}
\centerline{
\psfig{file=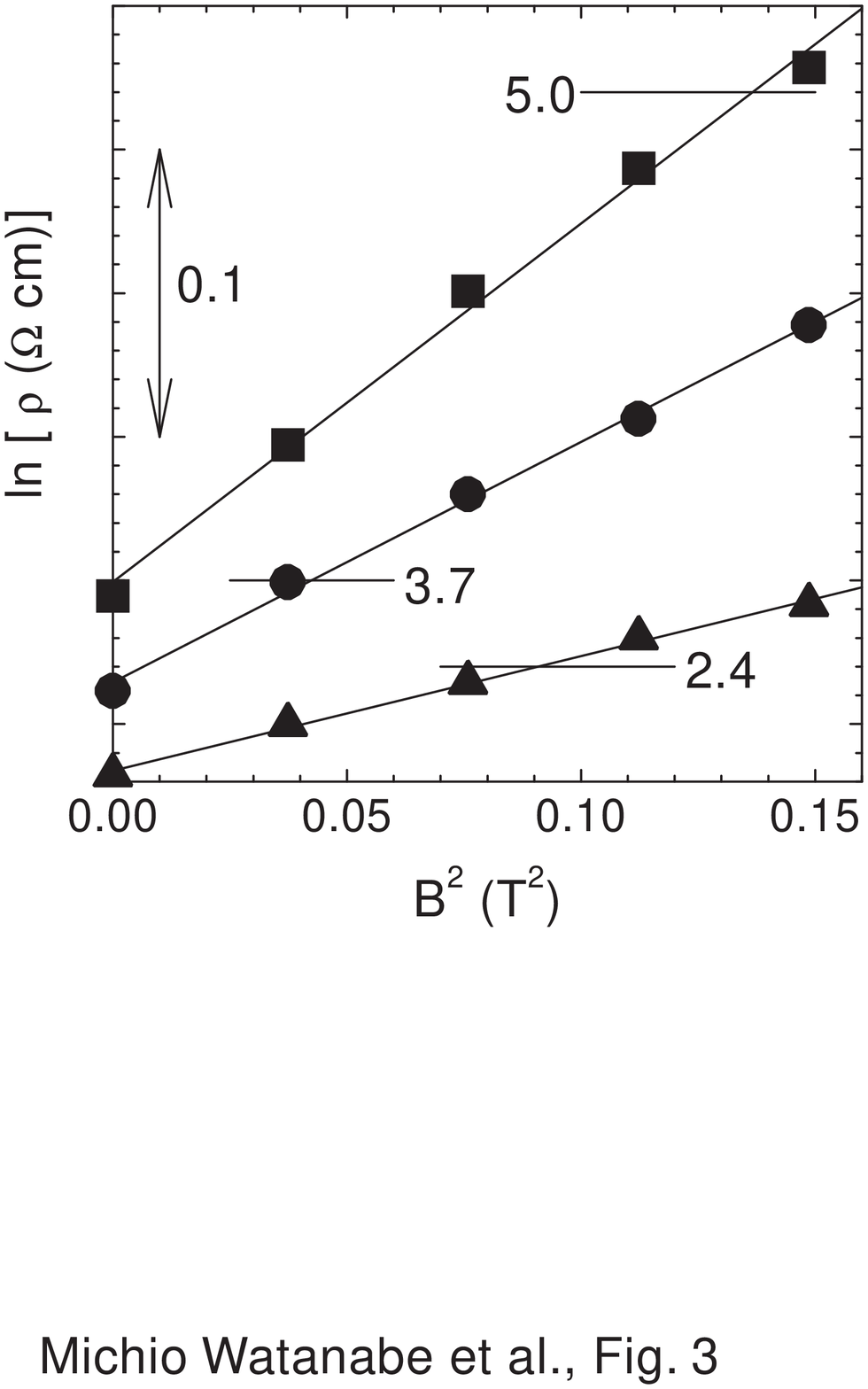,width=0.8\columnwidth,
bbllx=70,
bblly=255,
bburx=510,
bbury=725,clip=,
angle=0}
}
\caption{Logarithm of resistivity vs $B^2$ at constant 
temperatures for the sample having $N=1.840\times10^{17}$~cm$^{-3}$.   
From top to bottom the temperatures are 0.095~K, 0.135~K, and 0.215~K, 
respectively.  The solid lines represent the best fits.}  
\label{fig:B-dep}
\end{figure}
\begin{figure}
\centerline{
\psfig{file=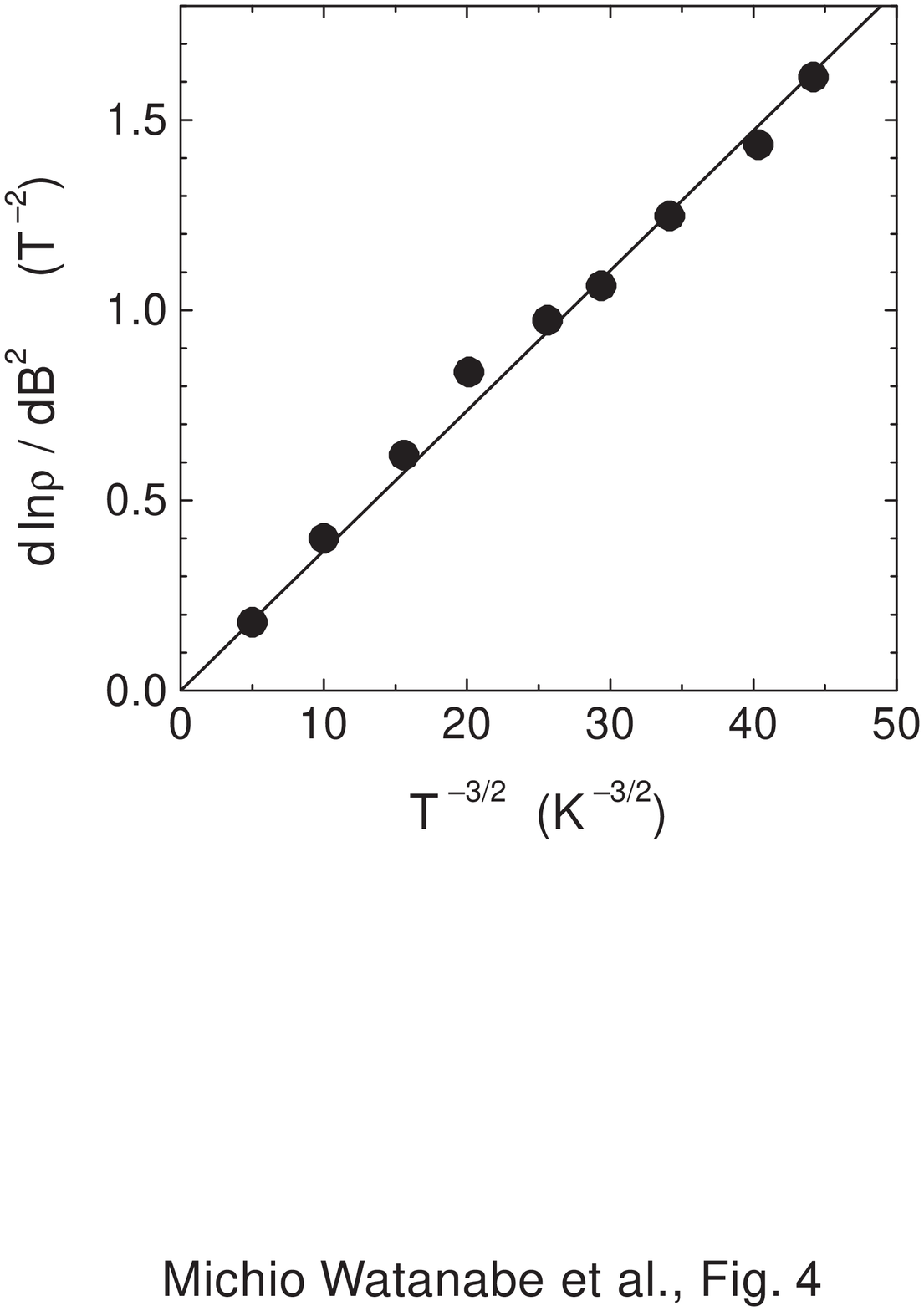,width=0.9\columnwidth,
bbllx=50,
bblly=285,
bburx=530,
bbury=725,clip=,
angle=0}
}
\caption{Slope $d\ln\rho/dB^2$ vs 
$T^{-3/2}$ for the sample having $N=1.840\times10^{17}$~cm$^{-3}$.  
The solid line represents the best fit.  } 
\label{fig:T32}
\end{figure}

%%%%%%%%%%%%%%%%%%%%%%%%%%%%%%%%%%%%
Our next step is to separate $T_0$ into $\epsilon$ and $\xi$.  
For $\xi/\lambda\ll1$, the magnetoresistance is expressed as 
\begin{equation}
\label{eq:rhoinB}
\ln[\rho(B,T)/\rho(0,T)]\approx 0.0015\,(\xi/\lambda)^4\,(T_0/T)^{3/2}, 
\end{equation} 
where $\lambda\equiv\sqrt{\hbar/eB}$ 
is the magnetic length in SI units.~\cite{Shk84}  
According to Eq.~(\ref{eq:rhoinB}), 
the magnetic-field variation of $\ln\rho$ at $T={\rm const.}$ 
is proportional to $B^2$, i.e., $\ln\rho(B,T) = \ln\rho(0,T) + C(T)B^2$, 
and the slope $C(T)$ is proportional to $T^{-3/2}$.  
In order to demonstrate that these relations hold 
for our samples, we show for the $N=0.989N_c$ 
sample $\ln\rho(B,T)$ vs $B^2$ in Fig.~\ref{fig:B-dep} 
and $C(T)$ determined by least-square fitting of $\partial\ln\rho/\partial B^2$ 
vs $T^{-3/2}$ in Fig.~\ref{fig:T32}.  
Since Eq.~(\ref{eq:rhoinB}) is equivalent to 
\begin{equation} 
\label{eq:gamma}
\gamma\equiv C(T)/T^{-3/2} 
\approx 0.0015\,(e/\hbar)^2\,\xi^{\,4}\,T_0^{\,3/2}\,, 
\end{equation} 
$\xi$ is given by
\begin{equation} 
\label{eq:xi}
\xi \approx 5.1\,(\hbar/e)^{1/2}\,\gamma^{1/4}\,T_0^{\,-3/8}\,.  
\end{equation}
In this way we have determined $\gamma$ for nine samples.  
The inset of Fig.~\ref{fig:chixi} shows $\gamma$ as a function of $T_0$.  
The value of $\gamma$ is almost independent of $T_0$, 
and if one assumes $\gamma \propto T_0^{\,\delta}$, 
one obtains a small value of $\delta = 0.094 \pm 0.005$ 
from least-square fitting.  
Figure~\ref{fig:chixi} shows $\xi$ and $\chi_{\rm imp}=\epsilon-\epsilon_h$ 
determined from Eqs.~(\ref{eq:T0}) and (\ref{eq:xi}).  
Here, $\epsilon_h$ is the dielectric constant of the host Ge, 
and hence, $\chi_{\rm imp}$ is the dielectric susceptibility 
of the Ga acceptors.  
We should note that both $\xi$ and $\chi_{\rm imp}$ are sufficiently larger 
than the Bohr radius (8~nm for Ge) and $\epsilon_h=15.4$ 
(Ref.~\ref{Cas80}), respectively. 
According to the theories of the MIT, both $\xi$ and $\chi_{\rm imp}$ 
diverge at $N_c$ as $\xi (N) \propto (1-N/N_c)^{-\nu}$
and $\chi_{\rm imp}(N)\propto(N_c/N-1)^{-\zeta}$, respectively.  
We find, however, both $\xi$ and $\chi_{\rm imp}$ do not show such 
simple dependencies on $N$ in the range shown in Fig.~\ref{fig:chixi}, 
and that there is a sharp change of both dependencies at $N\approx0.99N_c$.  
On both sides of the change in slope, the concentration dependence of 
$\xi$ and $\chi_{\rm imp}$ are expressed well by the scaling formula 
as shown in Fig.~\ref{fig:chixi}.  
Theoretically, the quantities should show the critical behavior 
when $N$ is very close to $N_c$.  So $\nu=1.2\pm0.3$ and $\zeta=2.3\pm0.6$ 
may be concluded from the data in $0.99<N/N_c$.  
However, the other region ($0.9<N/N_c<0.99$), where 
we obtain $\nu=0.33\pm0.03$ and $\zeta=0.62\pm0.05$, 
is also very close to $N_c$ in a conventional experimental sense.  
%%%%%%%%%%%%%%%%%%%%%%%%%%%%%%%%%%%%%%%%%%%%
%%%%%%  fig:T32  %%%%%%%%%%%%%%%%%%%%%%%%%%%
%%%%%%%%%%%%%%%%%%%%%%%%%%%%%%%%%%%%%%%%%%%%
%%%%%%%%%%%%%%%%%%%%%%%%%%%%%%%%%%%%%%%%%%%%

As a possible origin for the change in slope, we refer to the 
effect of compensation.  Although our samples are nominally 
uncompensated, doping compensation of less than 0.1\% 
may present due to residual isotopes that become $n$-type impurities 
after NTD.  In addition to the doping compensation, the effect known 
as ``self compensation" may play an important role near 
$N_c$.~\cite{Bha80}  
It is empirically known that the doping compensation 
affects the value of the critical exponents. 
Rentzsch {\em et al.} studied VRH conduction 
of {\em n}-type Ge in the concentration range of $0.2<N/N_c<0.91$, 
and showed that $T_0$ vanishes as $T_0\propto (1-N/N_c)^{\alpha}$ with 
$\alpha \approx 3$ for $K=38$\% and 54\%, where $K$ is the compensation 
ratio.~\cite{Ren98}  
Since $\alpha \approx \nu + \zeta$ [Eq.~(\ref{eq:T0})], we find for 
our $^{70}$Ge:Ga samples $\alpha=3.5\pm0.8$ for $0.99<N/N_c<1$ 
and $\alpha=0.95\pm0.08$ for $0.9<N/N_c<0.99$.  
Interestingly, $\alpha = 3.5\pm0.8$ agrees with $\alpha \approx 3$ found 
for compensated samples.  
Moreover, we have recently proposed the possibility that the conductivity 
critical exponent $\mu \approx 1$ in the same $^{70}$Ge:Ga only within 
the very vicinity of $N_c$ (up to about +0.1\% of $N_c$).~\cite{Ito99}  
An exponent of $\mu=0.50\pm0.04$, on the other hand, holds for a wide region of $N$ 
up to $1.4N_c$.~\cite{Wat98}  Again, $\mu \approx 1$ near $N_c$ may be viewed as the 
effect of compensation.  Therefore, it may be possible that the region of $N$ 
around $N_c$ where $\nu\approx1$ and $\mu\approx1$ 
changes its width as a function of the doping compensation.  
In the limit of zero compensation, 
the part which is characterized by 
$\nu\approx1$ and $\mu\approx1$ vanishes, i.e., 
we propose $\nu=0.33\pm0.03$, $\zeta=0.62\pm0.05$, and 
$\mu=0.50\pm0.04$ for truly uncompensated systems and that 
Wegner's scaling law of $\nu=\mu$ is not satisfied. 
In compensated systems, on the other hand, Wegner's law may hold 
as it does in the very vicinity of $N_c$.  The experiment on compensated 
Al$_x$Ga$_{1-x}$As:Si that showed $\zeta \approx 2$ 
and $\mu \approx 1$~(Ref.~\ref{Kat90}) is also consistent with the law. 
However, our preceding discussion needs to be proven in the future 
by experiments in samples with precisely and systematically controlled 
compensation ratios.   
%%%%%%%%%%%%%%%%%%%%%%%%%%%%%%%%%%%%%%
%%%%%  fig:chixi  %%%%%%%%%%%%%%%%%%%%
%%%%%%%%%%%%%%%%%%%%%%%%%%%%%%%%%%%%%%
%%%%%%%%%%%%%%%%%%%%%%%%%%%%%%%%%%%%%%

In summary, we have determined directly the localization length 
and the dielectric susceptibility arising 
from the impurities in nominally uncompensated NTD $^{70}$Ge:Ga 
samples near the critical point for the MIT. 
While the relation $2\nu\approx\zeta$ 
predicted by scaling theory~\cite{Kaw84,Imr81} holds for 
$0.9<N/N_c<1$, the critical exponents for localization length 
and impurity susceptibility change at $N/N_c\approx 0.99$.  
The small amount of doping compensation that is unavoidably present 
in our samples may be responsible for such a change in the exponents.  

We are thankful to T. Ohtsuki, 
B. I. Shklovskii, and M. P. Sarachik for valuable comments, 
and V. I. Ozhogin for the supply of the Ge 
isotope.  Most of the low-temperature measurements were carried out 
at the Cryogenic Center, the University of Tokyo.  
M. W. would like to thank the Japan Society for the Promotion of 
Science (JSPS) for financial support.  
The work at Keio 
was supported by a Grant-in-Aid for 
Scientific Research from the Ministry of Education, Science, Sports, 
and Culture, Japan.  The work at Berkeley was supported by the Director, 
Office of Energy Research, Office of 
Basic Energy Science, Materials Sciences Division of the U. S. 
Department of Energy under Contract No.~DE-AC03-76SF00098 and 
U. S. NSF Grant No.~DMR-97 32707.  

% now the references. delete or change fake bibitem. delete next three 
%   lines and directly read in your .bbl file if you use bibtex.

% figures follow here
%
% Here is an example of the general form of a figure:
% Fill in the caption in the braces of the \caption{} command. 
% Put the label that you will use with \ref{} command 
% in the braces of the \label{} command.
%
%
% tables follow here
%
% Here is an example of the general form of a table:
% Fill in the caption in the braces of the \caption{} command. 
% Put the label that you will use with \ref{} command 
% in the braces of the \label{} command.
% Insert the column specifiers (l, r, c, d, etc.) in the empty braces 
% of the \begin{tabular}{} command.
%
% \begin{table}
% \caption{}
% \label{}
% \begin{tabular}{}
% \end{tabular}
% \end{table}
%
%
%
\end{multicols}
\end{document}